\documentclass[10pt, conference, letterpaper]{IEEEtran}
\usepackage{graphicx}
\usepackage{epstopdf}
\usepackage{cite}
\usepackage{amsfonts}
\usepackage{amssymb}
\usepackage{amsmath}
\usepackage{mathrsfs}
\usepackage{bm}
\usepackage{amsmath}
\usepackage{mdwmath}
\usepackage{mdwtab}
\usepackage{url}
\usepackage[ruled,vlined,linesnumbered]{algorithm2e} 
\usepackage{subfigure}

\newtheorem{remark}{Remark}

\newenvironment{iarray}{\begin{IEEEeqnarray}{rCl}}{\end{IEEEeqnarray}\ignorespacesafterend}

\usepackage{makecell,multirow,diagbox}

\begin{document}

\title{Revealing Much While Saying Less: Predictive Wireless for Status Update
}
\author{\IEEEauthorblockN{Zhiyuan Jiang, Zixu Cao, Siyu Fu, Fei Peng, Shan Cao, Shunqing Zhang, and Shugong Xu}
	\IEEEauthorblockA{Shanghai Institute for Advanced Communication and Data Science, Shanghai University, Shanghai 200444, China.\\
	Emails: \{jiangzhiyuan, caozixu, fusiyu, pfly\_shmily, cshan, shunqing, shugong\}@shu.edu.cn
}
}
\maketitle

\begin{abstract}
Wireless communications for status update are becoming increasingly important, especially for machine-type control applications. Existing work has been mainly focused on Age of Information (AoI) optimizations. In this paper, a status-aware predictive wireless interface design, networking and implementation are presented which aim to minimize the status recovery error of a wireless networked system by leveraging online status model predictions. Two critical issues of predictive status update are addressed: practicality and usefulness. Link-level experiments on a Software-Defined-Radio (SDR) testbed are conducted and test results show that the proposed design can significantly reduce the number of wireless transmissions while maintaining a low status recovery error. A Status-aware Multi-Agent Reinforcement learning neTworking solution (SMART) is proposed to dynamically and autonomously control the transmit decisions of devices in an ad hoc network based on their individual statuses. System-level simulations of a multi dense platooning scenario are carried out on a road traffic simulator. Results show that the proposed schemes can greatly improve the platooning control performance in terms of the minimum safe distance between successive vehicles, in comparison with the AoI-optimized status-unaware and communication latency-optimized schemes---this demonstrates the usefulness of our proposed status update schemes in a real-world application. 

\end{abstract}

\section{Introduction}
\label{sec_intro}
Time-sensitive status update through a wireless interface is gaining more and more attention from researchers, with the emergence of Internet-of-Things (IoT) wherein most wireless networked devices (e.g., sensors, vehicles) are designed to communicate timely status information. A status update process represents an information flow from a source to a destination node (with possibly feedback) that is established to remotely estimate the status information generated at the source node. There are many fundamental differences between wireless design principles for status update and conventional packet-based systems. For instance, classical performance metrics such as throughput, communication latency and reliability cannot fully characterize a status update system---essentially, they are designed for packet-based systems which treat packets equally, whereas not suitable for status update systems wherein each individual packet is no longer the optimization target but the status flow. 

Towards this end, Age of Information (AoI) \cite{kaul12} has been proposed to characterize the flow-level performance of a status update process, which represents the time elapsed since the generation of the newest received status. Based on AoI \cite{kaul12}, a fundamental tradeoff is revealed between sampling frequency and communication delay which accounts for, e.g., networking scheduling delay and queuing delay---this tradeoff reflects on the received information timeliness. Intuitively, increasing the sampling frequency can reduce the information sampling error at the source, but in the meantime producing more packets that may lead to network congestion, i.e., increased communication delay. This tradeoff clearly consolidates the fact that status update no longer concerns transmissions for each packet---an increased sampling frequency in fact results in lower information density for each packet, in the sense that less packets contain meaningful status information. Consequently, an important system design implication is that the sensing (sampling) module has to be somewhat \emph{communication-aware}, e.g., if a network is congested, sources should be less frequently sampled (i.e., less packets) to reduce AoI \cite{sun17_tit}. However, while the decoupled sensing and communication modules enjoy simplified design, such a communication-aware sensing methodology complicates the system  implementation significantly, thus calling for a more practical solution. 

In this paper, we address the time-sensitive status update design in wireless networks systematically. The main contributions are summarized below.

	\textbf{Architecture:} A novel wireless communication protocol design is proposed specifically for status update, namely \emph{parallel communications}, which allows communication-agnostic sensing. Based on this architecture, status information is communicated by two parallel paths---one is the conventional Over-The-Air (OTA) wireless channels; the other is online learned, adaptive status model predictions. Specifically, the status packets are identified at the source node as those \emph{as expected} which are communicated by calibrated model predictions, and those \emph{unexpected} which are transmitted OTA. 

	\textbf{Networking:} A novel Status-aware Multi-Agent Reinforcement learning neTworking solution (SMART) is proposed to dynamically and autonomously adapt the transmit decisions of predictive wireless devices in an ad hoc network. Unlike conventional adaptive wireless transmissions, SMART accounts for the statuses of terminals and learned models, and makes transmit decisions accordingly. Essentially, the network performance is improved by parallel communications because the network load is alleviated by leveraging model predictions.

	\textbf{Testbed verification:} The proposed framework is implemented and the link-level performance is verified on a Software-Defined-Radio (SDR) testbed, with synthetic status data from a typical autonomous driving scenario and an LTE-based physical layer. It is shown that by adopting parallel communications, the channel occupancy is significantly reduced by online model predictions, while keeping the same status estimation error. Using Simulation of Urban MObility (SUMO), we test a wireless networked control application, i.e., multi dense platooning in steady state, and show that the proposed solution provides significantly better performance compared with current packet-based delay-optimized scheme, as well as the AoI-optimized status-unaware scheme.

The rest of the paper is organized as follows. In Section \ref{sec_exp}, a single-link example is shown to illustrate the core idea of parallel communications; the SDR verification follows immediately. Section \ref{sec_netw} describes the networking solution for parallel communications, i.e., SMART. Section \ref{sec_cs} provides a case study using SUMO, applying the proposed framework to multi dense platooning. We summarize the related work in Section \ref{sec_relatedWork}. Finally, in Section \ref{sec_cl}, conclusions are drawn with future directions discussed.
\begin{figure*}[!t]
	\centering    
	{\includegraphics[width=0.81\textwidth]{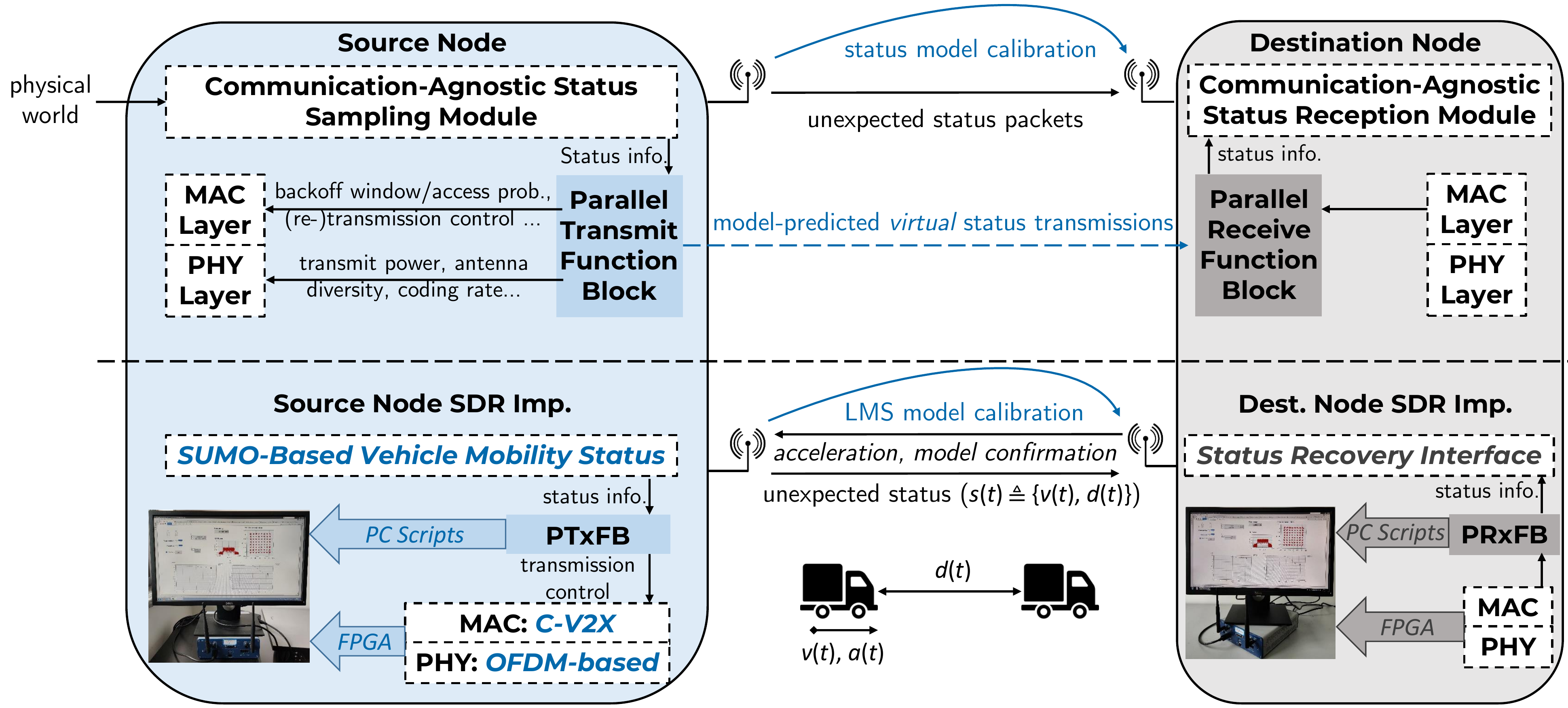}}
	\caption{Parallel communications diagram of a single link, and implementation on the SDR testbed.}
	\label{fig_pred}
\end{figure*}

\section{Predictive Wireless: A Single-Link Example and SDR Proof-of-Concept}
\label{sec_exp}
First, we would like to illustrate intuitively, plain and simple, on the central idea of predictive wireless, which is in analogous to human communications. Suppose that Alice and Bob live in different cities and Bob sends a letter to Alice everyday about whether it is rainy in his city on that day. This is a typical status update setting where a fixed update interval (one day) is adopted. However, further suppose that Bob's city rarely rains, and then he would soon discover that he could make the status update much more efficiently (in the sense of sending less reports), by defining and communicating a \emph{model} with Alice prescribing that whenever there is no letter, there is no rain; otherwise, Bob would report an \emph{model-unexpected status} indicating rainy. Obviously, the model in this case is a time-series forecasting model predicting no rain every time, and the status is a binary scalar value.

Such a simple, but useful idea can be generalized to improve the communication efficiency in status update-based wireless systems. In essence, such systems can benefit much from a status prediction functionality. The proposed system architecture diagram is shown in Fig. \ref{fig_pred}. We term this architecture as parallel communications, since the status information is communicated by both model predictions and OTA packets. The parallel communication operation is transparent to the upper layers or modules, i.e., communication-agnostic sensing module. Several key design aspects are illustrated in details as follows. We assume a time-slotted system and the time index is denoted by $t$ (the duration of a time slot is $1$ millisecond in the SDR implementation). At each time slot, a status (or a set of statuses) is sampled by the sensing module and fed into the Parallel Transmit Function Block (PTxFB).
\subsection{Model Estimation and Calibration}
A model, denoted by $\mathcal{M}_t(\cdot)$, is identified and estimated online by collecting the statuses over time at the source node, e.g., by well-known system identification and time-series forecasting techniques such as Auto-Regressive Integrated Moving Average (ARIMA) or Recurrent Neural Network (RNN)---in analogous to Bob discovering the no-rain model mentioned above. Specifically, the PTxFB at the source node takes input from the status sampling (sensing) module which samples a specific physical-world status with an arbitrarily high frequency (denoted by $f_{\mathsf{sample}}$) to minimize the sampling error. That is to say, the sensing module is totally regardless of the communication burden, i.e., communication-agnostic sensing. Based on the status input data, PTxFB fits a parametric model, which can be matrices in ARIMA or a neural network, to the status data. This model estimation process is performed online and constantly to adapt to status model changes in real world. Afterwards, the estimated model is transmitted to the destination node, i.e., calibrated online; in our design, we adopt a fixed model calibration interval, denoted by $T_{\mathsf{model}}$. It is essential that the source and destination maintain the same model and are synchronized---this issue is addressed in Section \ref{sec_issue} by a model confirmation mechanism.
\subsection{Parallel Transmitter}
At a source node, i.e., transmitter side, assuming that the PTxFB has a current status model which is calibrated with a corresponding destination node, a parallel transmitter works as follows.

	1. At time $t$, when a new status $\boldsymbol{s}(t)$ comes from the sensing module, the PTxFB makes a transmit decision:
	\begin{iarray}
	\mathsf{Transmit}(t) = \left\{\,
	\begin{IEEEeqnarraybox}[][c]{l?l}
		\IEEEstrut	
		\mathsf{True}, &\textrm{if } \operatorname{g}(\boldsymbol{s}(t),\bar{\boldsymbol{s}}(t)) > \delta; \\
		\mathsf{False},&\textrm{otherwise}, 
		\IEEEstrut
	\end{IEEEeqnarraybox}
	\right. 
	\end{iarray}
	where $\operatorname{g}(\cdot)$ denotes a error measure function, e.g., $\ell^1$, $\ell^2$ norms, the model-predicted status at time $t$ is denoted by $\bar{\boldsymbol{s}}(t)$ which is based on the estimated model and previous statuses, and $\delta$ denotes a threshold controlling how much status error the system can tolerate.
	
	2. The model prediction at time $t$ and the status estimation at the source and destination node (assuming calibrated) are respectively expressed as follows:
	\begin{iarray}
	\bar{\boldsymbol{s}}(t) &=& \mathcal{M}_t({\bar{\boldsymbol{s}}}(t-N_{\mathsf{input}}),\,\cdots,\,\bar{\boldsymbol{s}}(t-1)), \nonumber\\
	\hat{\boldsymbol{s}}(t) &=& \left\{\,
	\begin{IEEEeqnarraybox}[][c]{l?l}
	\IEEEstrut	
	\boldsymbol{s}(t), &\textrm{if $\mathsf{ACK}_t$} ; \\
	\bar{\boldsymbol{s}}(t),&\textrm{otherwise}, 
	\IEEEstrut
	\end{IEEEeqnarraybox}
	\right. 	
	\end{iarray}
	where $\mathsf{ACK}_t$ denotes a successful transmission at time $t$,\footnote{If the MAC layer protocol does not support acknowledgment feedback, each transmission is assumed to be successful.} and the input data size of the model is denoted by $N_{\mathsf{input}}$. 

Two points of explanation are in order. In the first step, essentially, when $\delta$ is larger, the status recovery error increases because only when the status variation compared with model prediction exceeds $\delta$ will there be a status packet transmission; otherwise, the error is smaller but the network occupancy is higher since more unexpected packets are transmitted OTA. In a distributed wireless network employing contention-based channel access mechanisms such as Carrier-Sense Multiple-Access (CSMA), higher occupancy directly leads to higher collision rate, and thus lower transmission reliability. Therefore, the threshold $\delta$ represents an important tradeoff that will be considered in more details in Section \ref{sec_netw}. Based on the second step, at the parallel transmitter, the status model takes the previous $N_{\mathsf{input}}$ estimated statuses as input, and produces the current model prediction, denoted by $\bar{\boldsymbol{s}}(t)$. Afterwards, $\bar{\boldsymbol{s}}^\prime(t)$ is compared with the real status ${\boldsymbol{s}}^\prime(t)$, and if the error is beyond the threshold, a status packet is transmitted. The estimated status at the destination node, considering status prediction output and OTA packets, is obtained at the source node, denoted as $\hat{\boldsymbol{s}}(t)$. The issue of unaligned estimations of statuses between source and destination are considered in Section \ref{sec_issue}.
\subsection{Parallel Receiver}
\label{sec_parR}
At a destination node, i.e., receiver side, again assuming the current status model is calibrated with a corresponding source node, the PRxFB works as follows.
\begin{iarray}
\hat{\boldsymbol{s}}^\prime(t) &=& \left\{\,
\begin{IEEEeqnarraybox}[][c]{l?l}
	\IEEEstrut	
	\boldsymbol{r}(t), &\textrm{if $\mathsf{ACK}_t$} ; \\
	\bar{\boldsymbol{s}}(t),&\textrm{otherwise}, 
	\IEEEstrut
\end{IEEEeqnarraybox}
\right. 	
\end{iarray}
where $\hat{\boldsymbol{s}}^\prime(t)$ denotes the status estimation at the destination node, and $\boldsymbol{r}(t)$ denotes the received status when successfully decoding a packet over the air interface, and hence $\boldsymbol{r}(t)=\boldsymbol{s}(t)$ neglecting sensing noise (in the testbed verification, sensing noise is added). Assuming calibrated status model, the model prediction at the source and destination should be aligned, i.e., when there is no packet transmitted, the destination node would use the model prediction as the received status and output to upper layers. Here, a successful reception is denoted by $\mathsf{ACK}_t$, which is the same with the transmitter side. 

\begin{remark}[\textbf{Transparency}] 
	Based on the above design, one of the main advantages of parallel communications is transparency to the sensing layer. Specifically, from the sensing layer's perspective, the status packets are constantly fed into the communication module without any consideration of the queuing or network load conditions. At the destination node, the status recovery module, which takes input from the parallel receiver, receives status packets at the same rate as the sampling rate at the source. Therefore, the upper layers of the status update system adopting parallel communications work completely transparent to the communication conditions; however, the true network occupancy is greatly reduced thanks to status model predictions. The term ``parallel'' refers to the fact that two communication paths are present: one is real OTA transmission and the other is model-prediction outputs by calibrated models between source and destination which also involve communications. 
\end{remark}

\subsection{SDR Implementation}
A two-vehicle platoon (a leader and a follower) with artificial status data is implemented on SDR
. The leader vehicle collects kinematic status information from the follower vehicle, and then assigns an acceleration to the follower vehicle to keep the distance between two vehicles a constant. The kinematic status information consists of the distance from the leader vehicle, and the instantaneous velocity and acceleration. 

The proposed scheme is implemented on an SDR platform consisting of two NI USRP-2974 \cite{usrp} devices. The USRP-2974 device is composed of of a Kintex-7 XC7K410T Field Programmable Gate Array (FPGA) baseband board and an embedded controller computer. National Instruments (NI) LabVIEW \cite{usrp},  a graphical programming tool for General-Purpose Processors (GPPs) and FPGAs, is utilized to develop the FPGA part and to run the higher layer control on the GPP. The physical layer of Cellular Vehicle-to-Everything (C-V2X) Mode 4 is implemented in the FPGA. A simplified MAC layer and the parallel communication framework are both implemented on GPP using LabVIEW. The details are specified below.

C-V2X Mode 4 supports vehicles to communicate with each other directly without base station coverage. In the PHY layer, one Resource Block (RB) is $180$~KHz ($12$ subcarriers of $15$ KHz subcarrier spacing). One subchannel is defined as a set of RBs in the same subframe, and the number of RBs can vary depending on applications. One subframe is $1$ ms. One vehicle User Equipment (UE) can transmit in one subchannel. A sensing-based Semi-Persistent Scheduling (SPS)---a decentralized resource allocation scheme---is adopted to enable direct V2X communications \cite{mol17}. The system parameters are as follows. The central frequency is $5.9$ GHz; the bandwidth is $20$ MHz ($100$ RBs) and is separated into $2$ subchannels; one single subchannel consists of $24$ RBs ($288$ subcarriers); transmit power is $-20$ dBm in our experiment. The modulation of control channel is QPSK and the coding scheme is tail biting convolution code; the modulation of data channel is 64QAM and the coding scheme is turbo coding. Both USRPs are settled on the table and the distance between the two is about $50$ centimeters.

In the experiment, two USRPs are utilized to mimic two vehicle UEs. The kinematic status information is generated by SUMO instead of real measurements considering the cost issue. However, the proposed scheme is tested on our SDR testbed, and hence the practicality is justified as long as the real status data is also predictable. In addition, Gaussian white nose is added to the velocity and the distance observations to simulate real-world imperfections, since kinematic sensors always suffer from measurement errors. Denote the observed noisy status information as $\bar{x}$ with $x$ being a real status.
The detailed procedure of our implemented parallel communication framework goes as follows. 

At the follower vehicle, every $1$ ms, it stores the current spacing ($\bar{d}(t)$), velocity ($\bar{v}(t)$) and acceleration ($\bar{a}(t)$) for model estimation and calibration. Denote $\bar{\boldsymbol{s}}(t) \triangleq [\bar{d}(t),\,\bar{v}(t),\,\bar{a}(t)]^{\mathsf{T}}$ and $\bar{\boldsymbol{w}}(t) \triangleq [\bar{d}(t),\,\bar{v}(t)]^{\mathsf{T}}$.
Every $100$ ms, the model parameters are calculated by the stored kinematic statuses by an Least Mean Squares (LMS) estimation method:
\begin{equation}
\label{lms}
\boldsymbol{A}(t) = [\bar{\boldsymbol{w}}(t),\,\cdots,\,\bar{\boldsymbol{w}}(t-99)]\cdot[\bar{\boldsymbol{s}}(t-1),\,\cdots,\,\bar{\boldsymbol{s}}(t-100)]^\dag,
\end{equation}
where $\boldsymbol{X}^\dag$ denotes the pseudo-inverse of $\boldsymbol{X}$ (in particular right pseudo-inverse). Meanwhile, the follower vehicle transmits the model parameters and piggyback the current kinematic status to the leader vehicle to calibrate the model. Every message conveying model parameters is time-stamped for model confirmation which is illustrated in the next paragraph.
Every $10$ ms, the status, i.e., the distance and velocity of the model prediction are compared with the actual observed status output from sensors. If the error (measured by $\ell^1$ norm) between the actual and predicted statuses is higher than a threshold ($0.1$ in our experiment), the actual status would be transmitted; otherwise no transmission and hence both ends would use the model prediction as communicated status at that time.

At the leader vehicle, every $1$ ms, it needs to transfer the status information to the higher layers. The parallel receiver procedure as described in Section \ref{sec_parR} is implemented. 
If the leader vehicle receives a model calibration message, it would substitute the previous model. A timestamp is received together with the message and is utilized for feedback as illustrated below.
The control algorithm at the leader vehicle takes the status information communicated by the proposed scheme as input, outputs an acceleration value based on \eqref{control} and transmits to the follower vehicle. This control information is transmitted based on conventional communication approach.
Every $10$ ms, the leader vehicle calculates the acceleration and transmits it to the follower vehicle together with the last timestamp received from the follower vehicle. This message is termed as the acceleration assignment message.

To calculate the acceleration value, the leader vehicle needs to know the desired distance, the current distance, the velocities of both vehicles and the acceleration of itself.
We adopt a specific Cooperative Adaptive Cruise Control (CACC) method to calculate the acceleration. The method is shown to be string-stable, i.e., a small turbulence in the platoon can be subsided by the control method \cite{vuk18}. Specifically,
\begin{iarray}
	\label{control}
a_{\mathsf{des},n} &=& \omega_1 (d_{\mathsf{des}}-\hat{d}_n) + \omega_2 (\hat{v}_n-\hat{v}_{n-1}) + \omega_3 (\hat{v}_n-\bar{v}_1) \nonumber\\
&& + \omega_4 a_{\mathsf{des},n-1} + \omega_5 a_{\mathsf{des},1},
\end{iarray}
where $d_{\mathsf{des}}$ is the desired inter-vehicle distance of the platoon which is limited by the safety requirements and should be kept as small as possible to reduce air drag, the status recovery of a status $x$ at the leader vehicle is denoted by $\hat{x}$ ($x$ being arbitrary), the desired acceleration of the $n$-th vehicle in the platoon (leader vehicle is the first) is denoted by $a_{\mathsf{des},n}$, and $\omega_i$, $i=1,...,5$ are constants satisfying certain conditions for string-stability\cite{vuk18}. In practice, the desired acceleration is fed into the lower controller of each vehicle such that $d_{\mathsf{des}}$ is maintained.
\begin{figure}[!t]
	\centering    
	{\includegraphics[width=0.45\textwidth]{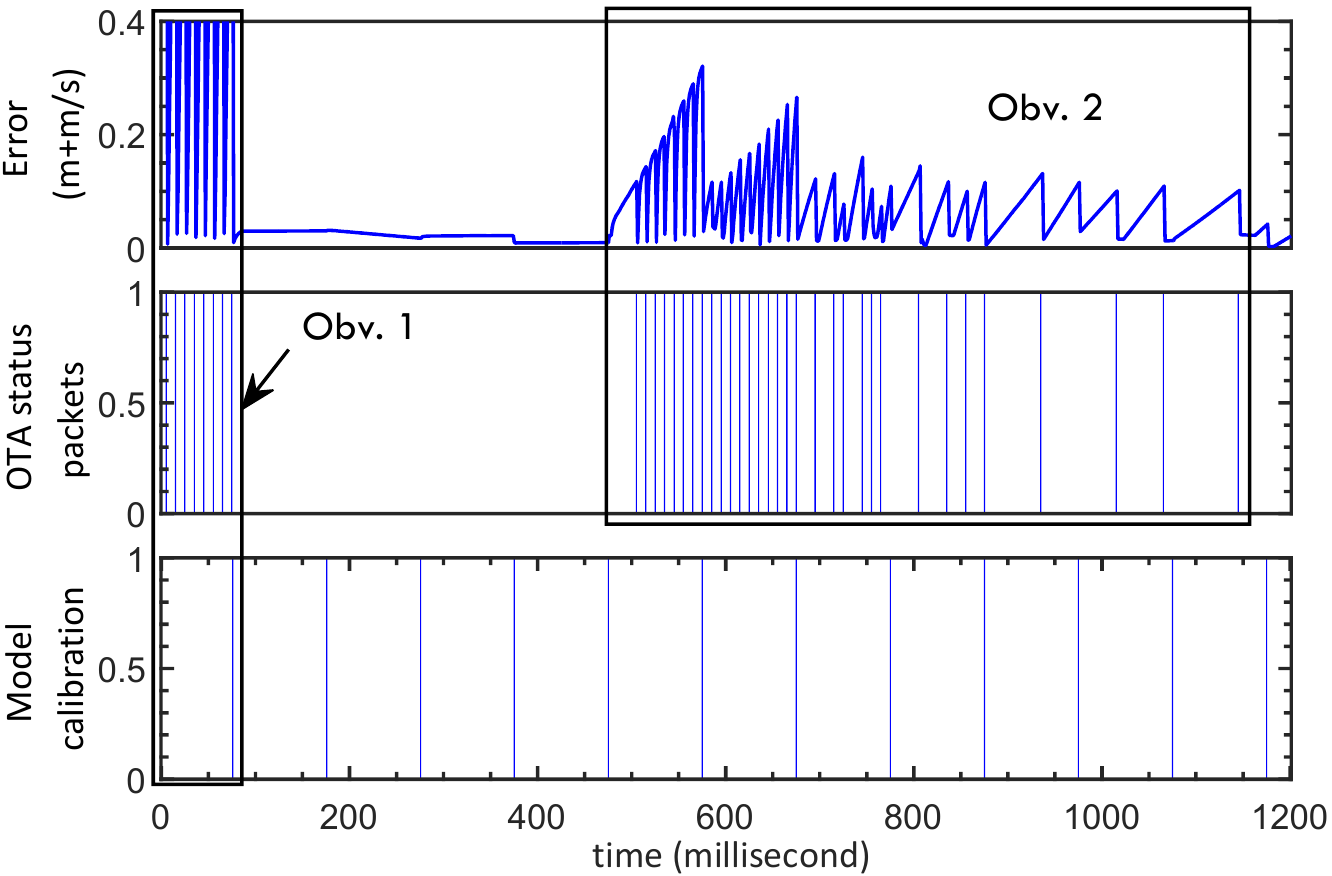}}
	\caption{Status recovery error measured on the SDR testbed. The transmission timing of both OTA status packets and the model calibration packets is also depicted.}
	\label{fig_sdr}
\end{figure}

In the feedback process at the leader vehicle, as mentioned above, a timestamp together with the model calibration message is first transmitted to the leader vehicle from the follower. Once this timestamp is received, the leader vehicle will transmit a confirmation packet back to the follower, in order to confirm the model calibration and calibration timing.
After the follower receives the timestamp the leader feeds back (a timeout mechanism is applied that after $10$ ms without feedback, the model calibration message is considered to be lost), it will compare the received timestamp with the last timestamp it transmits. If two timestamps are different, meaning that the leader vehicle fails to receive the model calibration message, the follower vehicle would not adopt the new model parameters; when the two timestamps match, the leader and follower vehicles both adopt the same and new model parameters.

Fig. \ref{fig_sdr} shows the status recovery error at the leader vehicle, tested on the SDR platform. The timing of status packets transmitted OTA and model calibration packets are also depicted for ease of exposition. The initial distance between two vehicles is $10$ meters and the initial velocity and acceleration of both vehicles are $10$ m/s and $0$ m/s${^2}$, respectively. The initial status models at both vehicles are generated randomly. The acceleration of the leader vehicle follows the formula:
\begin{iarray}
	\label{acc}
	a_1 &=& \left\{\,
	\begin{IEEEeqnarraybox}[][c]{l?l}
		\IEEEstrut	
		c(2478-t)(t-478), & \textrm{if }478\le t \le2478 ; \\
		0,&\textrm{otherwise}, 
		\IEEEstrut
	\end{IEEEeqnarraybox}
	\right. 	
\end{iarray}
where $c=4$ m/s$^2$. Two observations from Fig. \ref{fig_sdr} are in order. First (Obv. $1$ as denoted in Fig. \ref{fig_sdr}), before the first model status calibration, i.e., when the source and destination do not have an agreed model for status prediction, all packets are transmitted OTA; this is equivalent with the scheme without the proposed parallel communications. The status recovery error jumps to zero when there is an update and deviates in between two updates. After the first model calibration and confirmation, the status is predictable for about $400$ ms, i.e., error within the predefined threshold during this period, and hence no packets are transmitted OTA; however the status recovery error stays low due to model prediction outputs. Secondly (Obv. $2$), starting from $478$ ms, the leader vehicle begins to accelerate based on \eqref{acc}. During this period, the status model has changed due to acceleration and hence status recovery error starts to accumulate over time---once there is an packet transmission OTA, the error returns to zero. As we can see, the unexpected packet transmissions are dense during the beginning of this period. After about $200$ ms, the model is calibrated based on online estimations to capture the model change, and hence the status error is small afterwards with relatively less frequent unexpected packet transmissions. Overall, we can conclude that the proposed scheme is feasible in practice, and that parallel communications can significantly reduce the packet transmissions OTA and maintains a low status recovery overhead.

\subsection{Issues and Solutions}
\label{sec_issue}
Several practical issues found in experiments are discussed here.
\emph{Status misalignment}: One of the crucial requirements for the proposed design is that the status estimations (including model predictions) should be aligned at both ends, i.e., $\hat{\boldsymbol{s}}(t) = \hat{\boldsymbol{s}}^\prime(t)$, $\forall t$. Such a requirement could be jeopardized by the following causes.
 
	1) Model calibration packet transmission error: The packet that contains the newly estimated model using recent status data is lost due to channel error, e.g., collisions or channel fading. 
	
	2) OTA status packet transmission error: The transmission of an unexpected status packet OTA is erroneous. 
	
	3) Timing misalignment: Due to hardware signal processing latency or other causes, the timing of a status packet received from the air interface at destination could be unaligned with the source. For example, the transmitter sends a packet at time $t_1$, but the perceived sending time is $t_1+\tau$ where $\tau$ denotes the processing latency. This issue is identified in the testbed and is believed to be ubiquitous in practice.

Notice that status misalignment by even one packet loss causes persistent status recovery error over time. To see this, assuming that one model calibration packet is lost, all subsequent model predictions would be different, due to the fact that the source node would use the newly calibrated model but the destination node would still use the old model. The situation with status packet loss is similar, for that new model predictions would be unaligned due to different views on status history at source and destination. Timing misalignment also causes the same problem. 

To address this issue, we adopt three methods that proven effective to counteract the three causes in the implementation, namely model confirmation, periodical model and status retransmissions, and timestamps. There three methods are used jointly. The model confirmation packet is a feedback packet transmitted by the destination node immediately ($1$ timeslot to allow signals to be processed) after it receives a model calibration packet, containing a timestamp indicating the time when the destination receives the model calibration. If the source receives the confirmation packet, in which case the source and destination have agreed on the new model and the time to begin using it, then the issue is solved. In practice, we use repetition to ensure the successful reception of the model confirmation packet since it is vital. Note that repetition cannot be directly applied to the original model calibration packet since the destination would be confused about the exact timing to use the new model. Moreover, periodical packets which contain the model calibration and current status are transmitted, termed as correction packets, to avoid the following possible problem: When an unexpected status packet is lost, long time would pass without any unexpected status packet being transmitted if the model prediction was precise, in which case the model predictions at source and destination would be unaligned for a long time, resulting in severe status recovery error. In practice, the time interval between correction packets is long, e.g., $1$ second, to reduce the overhead thereby. Last but not least, a useful method to counteract timing misalignment brought by hardware impairments is piggybacking the timestamp of the status packet that is transmitted OTA. By doing this, the source and destination nodes can agree on the correct status timing that is carried by the timestamp and thus generate aligned status predictions.

\begin{algorithm}
	\caption{SMART}
	\label{alg:SMART}
	\small
	\textbf{Phase 1: Election of Supervision Node}\\
	Every destination node is elected as a Supervision Node (SN).\\	\textbf{Phase 2: Offline Single-Agent Training} \\
	\textbf{Initialization}: 
	Source nodes: Initialize their DDPG by the normal distribution.\\
	\label{p}
	\For{$n=1:N$}{		
		\For{$m=[m_{\mathsf{min}}:m_\mathsf{int}:m_{\mathsf{max}}]$}
		{
			DDPG training for source node-$n$ to solve the MDP expressed in \eqref{c2go} with given $m_n=m$. Afterwards, save the DDPG parameters ($\boldsymbol{w}_{m,n}$) to its database.
		}	
	}
	\label{store}
	\textbf{Phase 3: Online Auxiliary Cost Adaptation} \\
	\textbf{Initialization}: 
	Source nodes: Initialize their DDPG by $\boldsymbol{w}_{m_\mathsf{max},n}$ ($n=1,...,N$). SNs: Assign $m=m_\mathsf{max}$ as the initial auxiliary cost.\\
	\For{$t = 1:T$}{
		\For{$n=1:N$}{
			\If{The output of DDPG of source node-$n$ is $\mathsf{transmit}$ based on its state}{
				Source node-$n$ transmits, following the underlying MAC protocol. 
			}
			\Else{Source node-$n$ stays silent.}
		}	
		\If{$t\%\mathsf{evaInt}=0$}
		{\label{interval}
			$\mathsf{cost}=\sum_{\tau=t-\mathsf{evaInt}+1}^{t} \mathcal{E}(\boldsymbol{x}_1(\tau),...,\boldsymbol{x}_N(\tau))$\\
			\If{$\mathsf{cost}-\mathsf{costPrev} > \delta$ and the number of transmission collisions increases}{
				$m = \max\{m + m_\mathsf{int},\,m_{\mathsf{max}}\}$ at the corresponding source nodes.
			}
			\label{adapt}
			\ElseIf{$|\mathsf{cost}-\mathsf{costPrev}|<\delta$}{$m$ is unchanged.}
			\Else{
				$m = \min\{m - m_\mathsf{int},\,m_{\mathsf{min}}\}$ at the corresponding source nodes.
			}
			$\mathsf{costPrev} = \mathsf{cost}$
		}
	}
\end{algorithm}

\section{Networking of Predictive Wireless Devices for Status Update}
\label{sec_netw}
In this section, we describe in general a networking scheme for efficient communications among ad hoc wireless devices running the parallel communication protocol. One of the critical issues of this networking scheme that distinguishes from conventional ad hoc network problems is that how to properly make transmit decisions when the decisions depend on locally-observable \emph{status} information, i.e., status-aware. The scheme is inspired by the Whittle's index methodology \cite{web90} and is explained as follows. 

Denote $\boldsymbol{x}_n(t) \triangleq [\boldsymbol{s}_n(t),\,\hat{\boldsymbol{s}}_n(t)]$ as the Markov state for $n$-th source node in the system at time $t$, wherein $\boldsymbol{s}_n$ denotes the local status information, and $\hat{\boldsymbol{s}}_n(t)$ denotes the status recovery at the $n$-th destination (estimated at source nodes based on aligned status). Note that for non-Markov status evolution, a collection of finite length history of statuses can be approximately defined as Markov states which is common in practice. The overall system space is hence $\{\boldsymbol{x}_n(t)|n=1,...,N\}$ wherein $N$ is the total number of source nodes. Denote $\boldsymbol{u}(t) = \{u_n(t)|n=1,...,N\}$ as the transmit decision of sources at time $t$; $u(t)=1$ denotes transmit and silent otherwise. The Markov Decision Process (MDP) can be defined as
\begin{flalign}
\label{MDP1}
\textbf{MDP-1:}&&\mathop{\textrm{minimize}}\limits_{\boldsymbol{u}(t)}  \,\,&  \limsup_{T \to \infty} \frac{1}{T}\sum_{t=0}^{T-1} \mathcal{E}(\boldsymbol{x}_1(t),...,\boldsymbol{x}_N(t)) &&\nonumber\\
&&\textrm{s.t.,}\,\,  & \textrm{source-$n$ knows }\boldsymbol{x}_n(t),\,\sum_{n=1}^N u_n(t)=1,&& \nonumber
\end{flalign}
where $\mathcal{E}(\cdot)$ denotes a predefined error function. MDP-1 is essentially a Partially Observable MDP (POMDP) as seen by each source-destination pair, since each source can only observe its local current status. MDP-1 suffers from the curse of dimensionality when using e.g., value iterations. In addition, MDP-1 is a POMDP which does not have a general solution method---multi-agent reinforcement learning techniques may be leveraged but with convergence issues \cite{pes00}. 

Towards this end, we adopt the concept of Whittle's index \cite{web90} to decouple MDP-1. The idea of this approach is that instead of considering all nodes simultaneously, the problem is decoupled and reduced to transmit decisions for one source node. To avoid the trivial, selfish and useless solution of always transmitting, each transmission is associated with a \emph{auxiliary cost} of $m_n$. In other words, the decoupled MDP problem for source node-$n$ is formulated as
\begin{equation}
\label{c2go}
f(\boldsymbol{x}_n) + \hat{J}^*  =  \min \left\{ \begin{array}{l}
\mathcal{E}^{(0)}_{\boldsymbol{x}_n}  + \sum_{\boldsymbol{x}_n^{\prime}} \mathcal{P}^{(0)}_{\boldsymbol{x}_n \boldsymbol{x}_n^{\prime}} f(\boldsymbol{x}_n^{\prime}) ,\\
m_n +  \mathcal{E}^{(1)}_{\boldsymbol{x}_n} +\sum_{\boldsymbol{x}_n^{\prime}} \mathcal{P}^{(1)}_{\boldsymbol{x}_n \boldsymbol{x}_n^{\prime}} f(\boldsymbol{x}_n^{\prime})
\end{array} \right\}, 
\end{equation}
wherein the top and bottom terms in the minimization operator represent the cost-to-go from state $\boldsymbol{x}_n$ onwards with the action of silent and transmit, respectively. The expected cost functions are denoted by $\mathcal{E}^{(0)}_{\boldsymbol{x}_n}$ and $\mathcal{E}^{(1)}_{\boldsymbol{x}_n}$ respectively for both actions (assuming the error function is decomposable, e.g., $\ell^1$, $\ell^2$ norms); the transition matrices $\mathcal{P}^{(0)}_{\boldsymbol{x}_n}$ and $\mathcal{P}^{(1)}_{\boldsymbol{x}_n}$ are denoted likewise. The relative cost-to-go function of state $\boldsymbol{x}_n$ and the average cost are denoted by $f(\boldsymbol{x}_n)$ and $\hat{J}^*$ respectively. It has been shown that the solution solving \eqref{c2go} for each source node and selecting the node with the largest index ($\max\{m_n|n=1,...,N\}$) to transmit leads to near-optimal performance in many problems \cite{kadota18,hsu18,jiang18_itc}. Inspired by these results, we design the following scheme, as shown in Algorithm \ref{alg:SMART}, which is termed as SMART. Due to the fact that the status evolution is unknown, as well as that no supervisor is available in real world, we adopt a deep reinforcement learning-based approach. The status in general is comprehended using a Deep Neural Network (DNN), specifically a Deep Deterministic Policy Gradient (DDPG) approach \cite{ddpg}, while in practice if the status is straightforward, simpler approaches can be adopted.

In Algorithm \ref{alg:SMART}, we use $m_{\mathsf{max}}$ and $m_{\mathsf{min}}$ to denote the maximum and minimum auxiliary costs, respectively. In practice, $m_{\mathsf{max}}$ and $m_{\mathsf{min}}$ are hyper-parameters that depend on network conditions to provide adaptation capabilities inside the interval $[m_{\mathsf{min}}, m_{\mathsf{max}}]$. SMART works roughly as follows. In the single-agent training phase, a set of possible auxiliary cost $m$ is trained for each source node as if only itself is updating with the additional auxiliary cost. A mapping from each $m$ to the DDPG parameters is stored in the node's database. Afterwards, in the auxiliary cost adaptation phase, the SNs\footnote{In practice, it is found that representative (not all) destination nodes elected as SN are sufficient.} observe the network conditions (i.e., collisions, channel idle and status values) for a period of time $\mathsf{evaInt}$ in Step \ref{interval} and thereby adjust the auxiliary cost values. When the cost increases due to congestion, the auxiliary cost should be increased to discourage source nodes from transmitting; in other cases, the auxiliary cost should be decreased, or stays the same after convergence. The corresponding source nodes switch their DDPG parameters based on their respective SNs' feedback. SMART is also compatible with different MAC protocols. For example, nodes would transmit randomly based on the backoff window size in CSMA and C-V2X protocols. Eventually, the network converges to a situation wherein appropriate auxiliary costs are attained such that only nodes with urgent transmission needs (e.g., large status prediction error) are transmitting and the network is properly loaded. Note that the auxiliary costs can be different among nodes since each source node follows its corresponding SN's feedback. In the following section, we will observe in details how a system of platooning vehicles behave under SMART and parallel communications.
\begin{figure}[!t]
\centering    
{\includegraphics[width=0.5\textwidth]{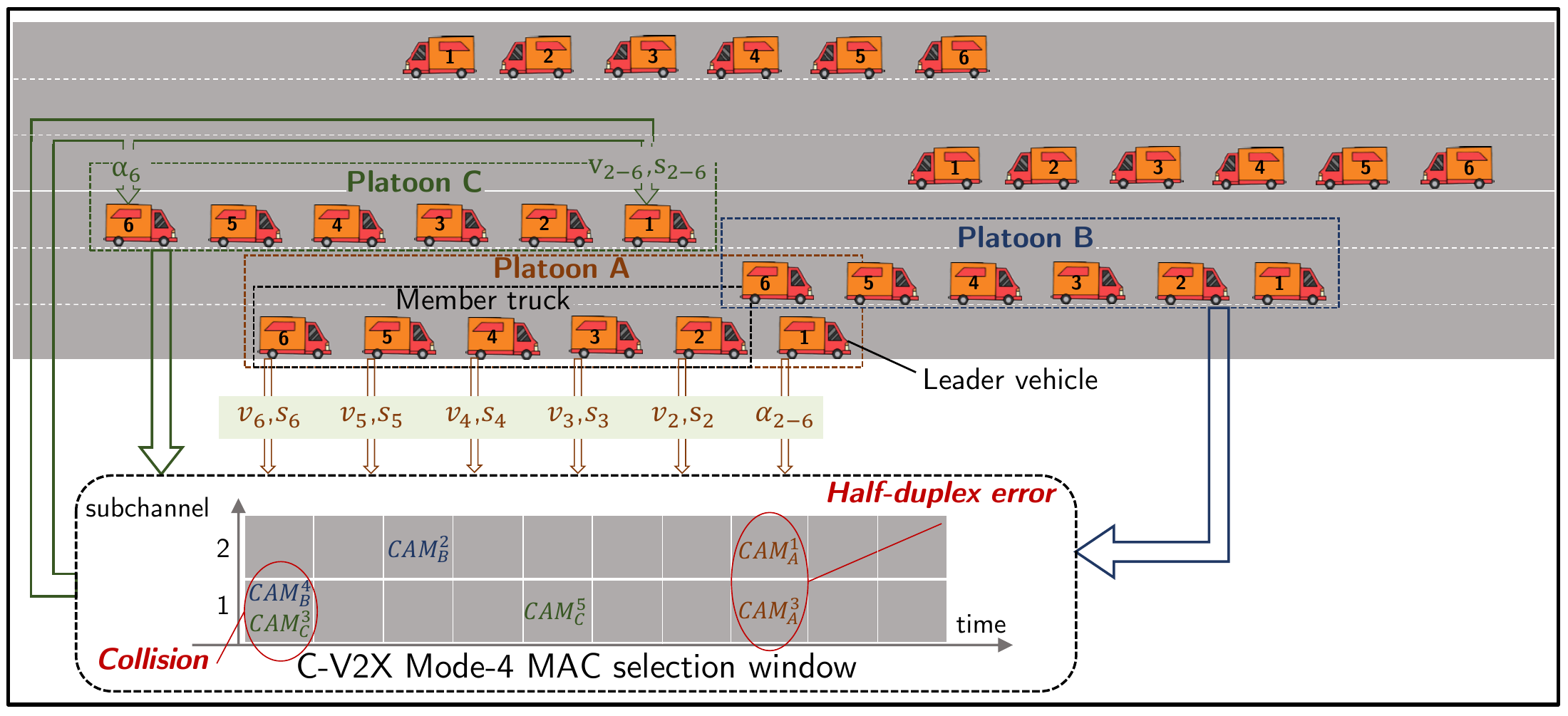}}
\caption{A multi dense platooning scenario wherein each leader vehicle controls the corresponding platoon. The C-V2X mode-4 MAC layer is implemented on SUMO to simulate vehicular communications.}
\label{fig_platoon}
\end{figure}
\begin{figure*}[!t]
	\centering    
	{\includegraphics[width=1.0\textwidth]{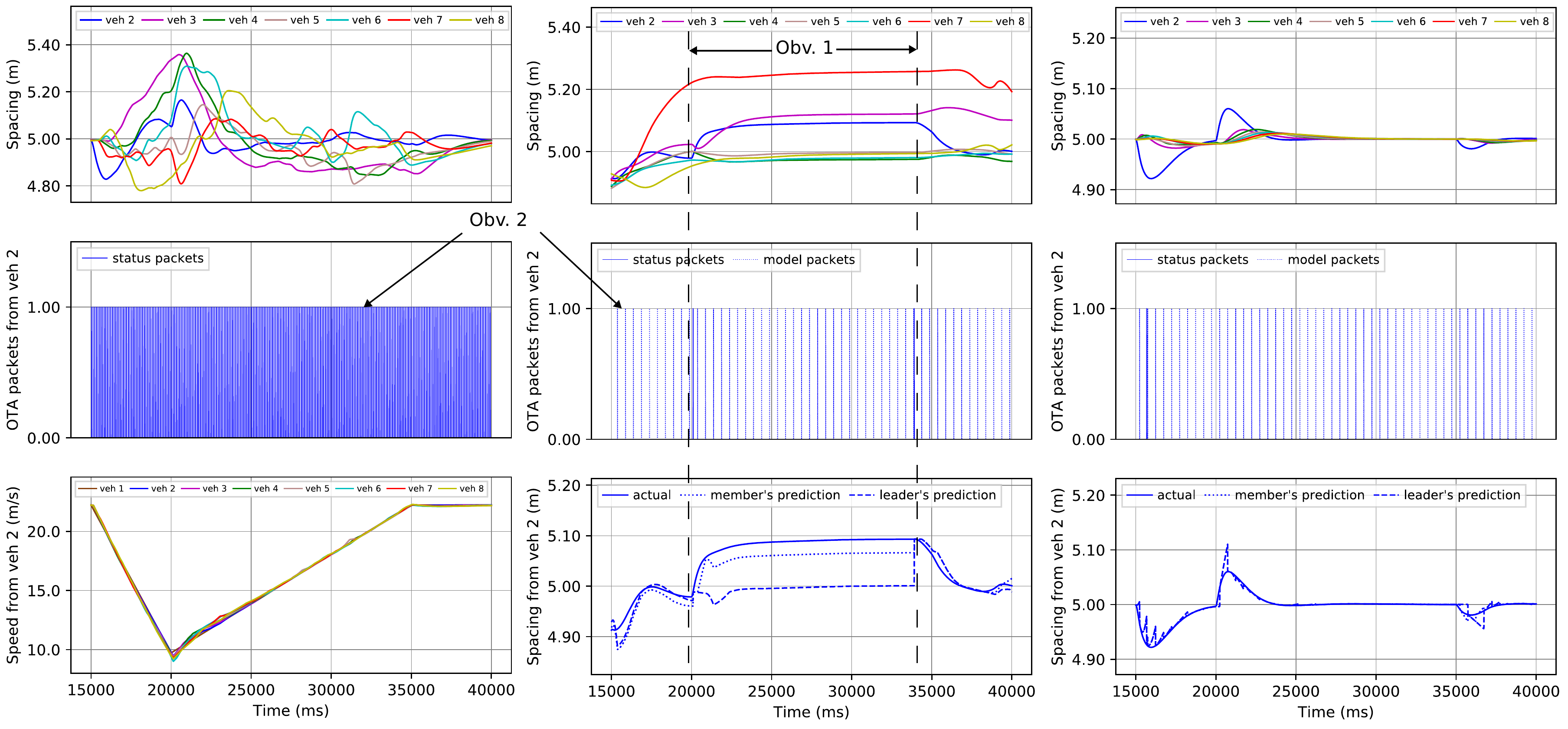}}
	\caption{Performance comparisons with status-unaware scheme with the optimal update interval. The left, middle and right columns represent optimal status-unaware, parallel communication without and with correction packets, respectively. The top, middle and bottom rows represent vehicle distances from the front vehicles, status and correction packets transmissions, and status predictions at both ends. The bottom-left figure shows the velocity of the leader vehicle. }
	\label{fig_sumoRes}
\end{figure*}
\begin{figure}[!t]
	\centering    
	{\includegraphics[width=0.4\textwidth]{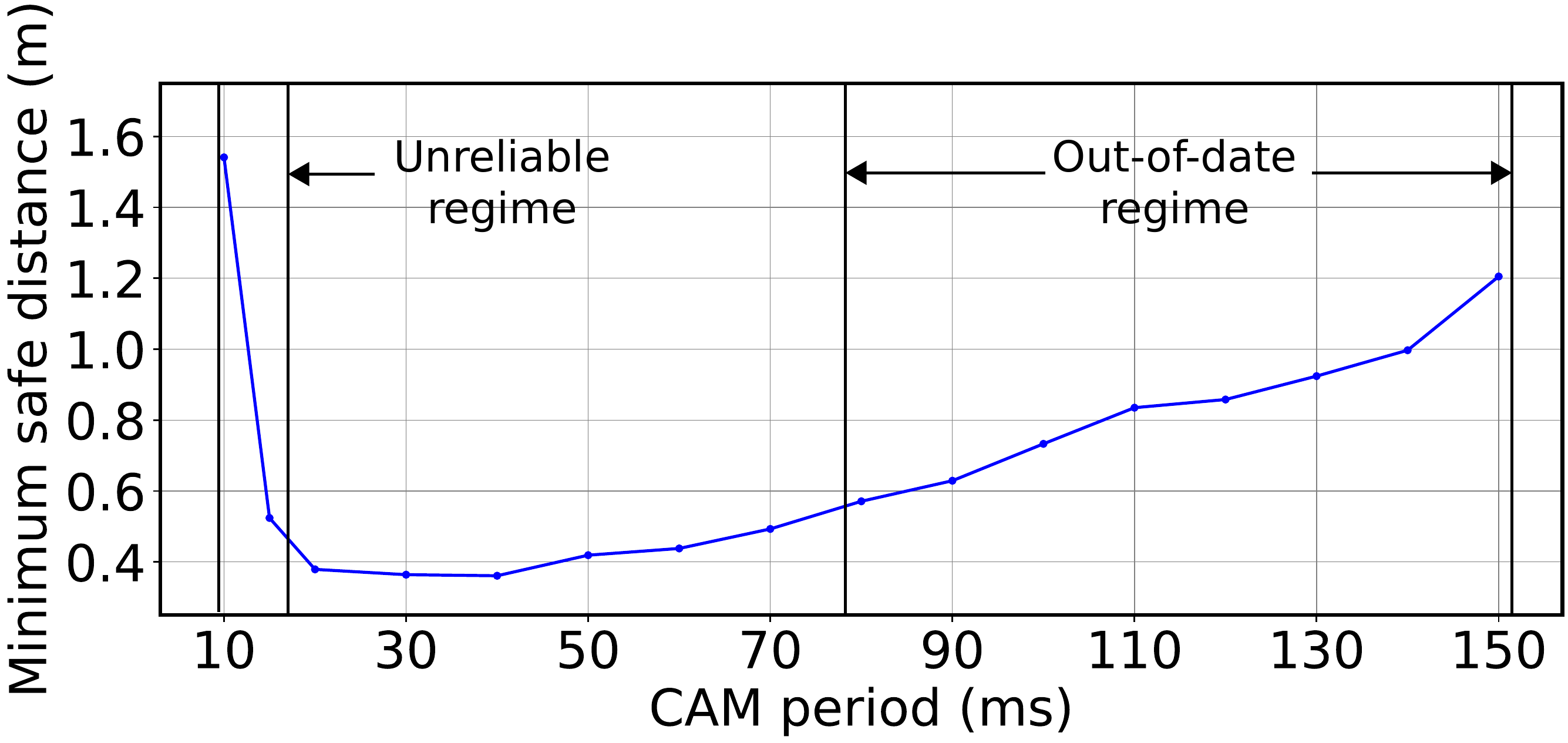}}
	\caption{Minimum safe distance versus the status update interval by status-unaware update schemes.}
	\label{fig_aoi}
\end{figure}
\begin{figure}[!t]
	\centering    
	{\includegraphics[width=0.43\textwidth]{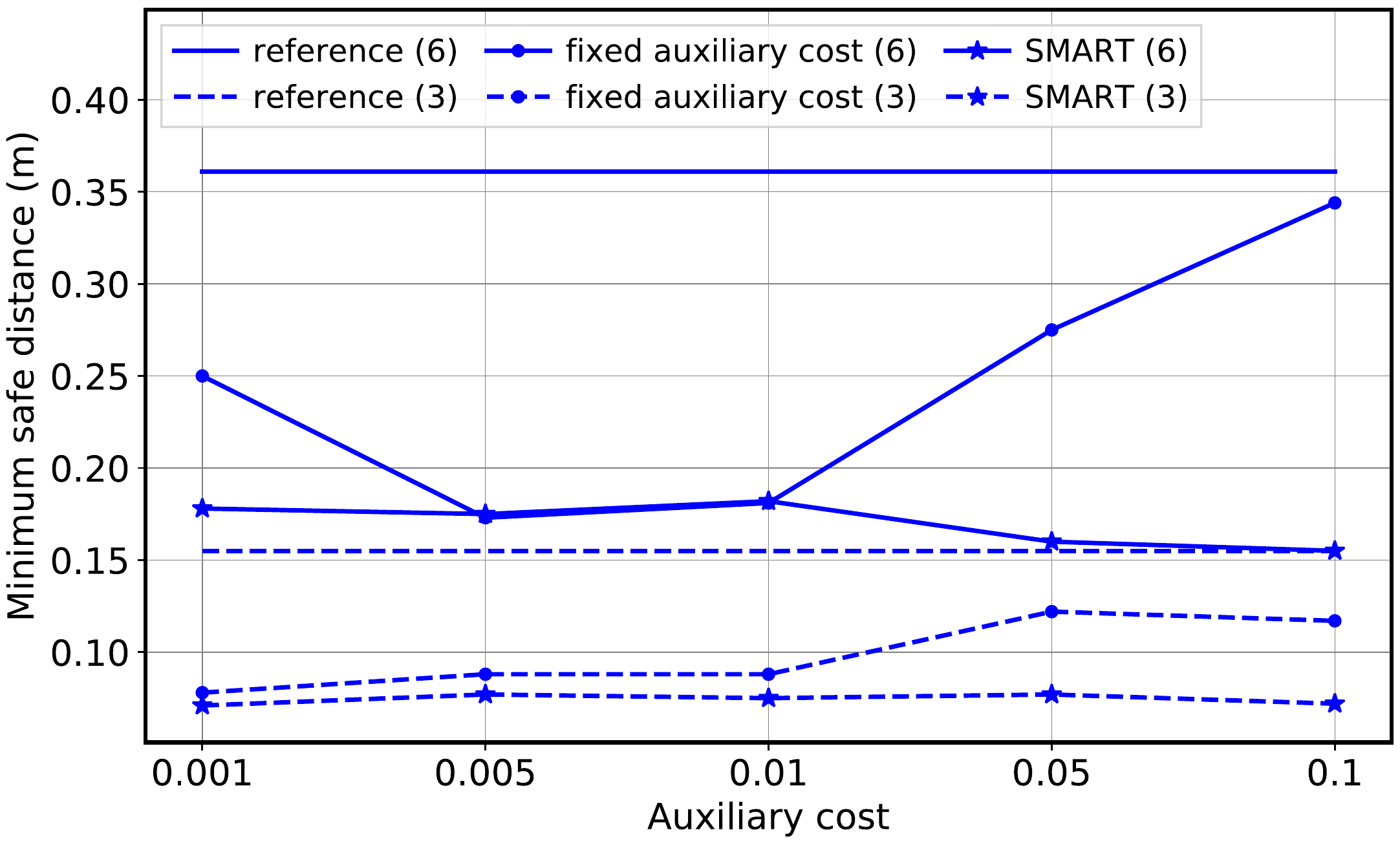}}
	\caption{Effectiveness of auxiliary cost adaptation in SMART. The number in the bracket represents platoon quantity on the road (each with $8$ vehicles).}
	\label{fig_thres}
\end{figure}
\section{Case Study: Multi Dense Platooning in Steady State}
\label{sec_cs}
In this section, we demonstrate through SUMO a Wireless Networked Control (WNC) application, namely multiple dense platooning in steady state using C-V2X communication protocol and the proposed framework (see Fig. \ref{fig_platoon}). The application is selected for two reasons. First, one of the most promising 5G vertical applications is C-V2X-enabled autonomous driving and platooning is perhaps the most attractive fully autonomous driving technology; secondly, platooning represents a WNC system wherein timely (milliseconds level) and precise status update and control are essential. 

\textbf{Scenario and Platoon Control}: Several platoons, each led by a leader vehicle, travel close to each other (within the communication range of C-V2X which is typically $700$ meters \cite{v2x}) on the highway in steady state, i.e., after platoons are formed. We consider only the longitudinal drive control, that is the vehicles travel on a straight line. In each platoon, the leader vehicle is driven by human and the others are controlled by the leader vehicle based on control algorithms such as \eqref{control} and wireless signals. Naturally, for the leader vehicles to make precise control, the status information it collects from its follower vehicles, including distances from the front vehicle and instantaneous velocities, should be as timely as possible. The ultimate goal of the system is to save fuel while ensuring safety, and hence the vehicles are designed to follow its corresponding front vehicles as close as possible while not crashing into them. In the simulations, we set a predefined desired distance between two vehicles and let the control algorithm to maintain this desired distance---the minimum safe distance is therefore the maximum distance reduction from the desired distance during the whole trajectory. The simulation scenario is a two-way, four-lane highway with a length of $1500$ meters. The length of vehicles is $5$ meters. The leader vehicle enters with a speed of $10$ m/s, then accelerates to $22.2$ m/s in $5$ seconds; from the $15$-th second, the leader decelerates to $9.7$ m/s in $5$ seconds, and then accelerates to $22.2$ m/s in $15$ seconds (depicted in the bottom-left figure in Fig. \ref{fig_sumoRes}). The acceleration of vehicles is restricted to $[-2.94,4]$ m/s$^2$ to avoid very rapid and abrupt changes in speed. The simulation time step is $1$ ms. The actuation delay and sensing delay are ignored to focus on information timeliness, and hence the performance evaluation can be viewed as an optimistic bound.

\textbf{Network Protocol}: The underlying MAC- and PHY-layer protocols that convey the status and control information are based on C-V2X Mode 4 \cite{mol17}. Specifically, as shown in Fig. \ref{fig_platoon}, an SPS decentralized time/frequency resource allocation scheme is adopted wherein a vehicle UE with packet (Cooperative Awareness Message, CAM) to send could choose uniformly randomly from a pool of time/frequency resources. The pool is composed of several subchannels in the frequency domain and a number of subframes with length denoted by Resource Reservation Interval (RRI). Thereby, as the selection is fully autonomous, collisions may happen especially when the system is heavily loaded. In addition to collisions, Half-Duplex (HD) error also occurs in C-V2X Mode 4, which represents the error caused by receiving a packet while transmitting on the same subframe. In the simulations, we set the RRI to be $10$ ms and the number of subchannels to be $2$, in consistency with current standards \cite{mode4}. 

\textbf{Parallel Communications Implementation}: The majority of implementation details are the same as described in Section \ref{sec_exp} and \ref{sec_netw}. In particular, the status model is estimated based on a sliding-window LMS method \eqref{lms}. We choose the cost function $\mathcal{E}(\cdot)$ in SMART as the standard deviation from the desired distance between successive vehicles. The model calibration time interval is $500$ ms. 

\textbf{Simulation Results}: In Fig. \ref{fig_sumoRes}, the proposed parallel communication framework is tested in comparison with the status-unaware scheme using the optimal status update interval. Status-unaware schemes represent those without knowledge of the status content. Therefore, the best they can achieve is to optimize over the update interval; we obtain the optimum by simulating over a set of status update intervals from $10$ ms to $150$ ms, with step size of $10$ ms, and the optimum is shown to be $40$ ms (see Fig. \ref{fig_aoi}). Note that optimizing the average AoI among all vehicles is equivalent to optimizing over update intervals since the statuses are generate-at-will \cite{kadota18}. Intuitively, the tradeoff is that a smaller update interval leads to worse congestion, but lower update waiting delay; vice versa. Comparing among the first column in Fig. \ref{fig_sumoRes}, it is observed that the parallel communication scheme with correction packets outperforms the others significantly, showing that by using model predictions, the network load is reduced such that the status attained and control is more timely and hence the vehicle distance variance is much smaller. Observing the middle column (Obv. 1), it is shown that without correction packets (which are transmitted periodically and carry status information) the parallel communication scheme suffers from the issue we discussed in Section \ref{sec_issue}. That is, after a status packet loss, the source and destination are unaligned as can be observed in the bottom-middle figure; hence, the control is affected and the vehicle distance is kept to be larger than desired until the next unexpected status packet transmission. The transmissions of correction packets certainly entail overhead. From Obv. 2, it is shown that the optimal status packet update interval for status-unaware schemes is $40$ ms, and the correction packet transmission interval (same with model calibration packets) is $500$ ms. Therefore, the additional overhead of correction packets is well worth it since it is relatively small and useful. 
A key observation in Fig. \ref{fig_aoi}, which shows that the optimal status update interval of status-unaware schemes is about $40$ ms, is that in status update, sometimes it is better to be timely but unreliable, than ultra-reliable but sacrificing timeliness. This is different with the current ultra-Reliable Low-Latency Communications (uRLLC) principle. Specifically, as shown in the figure, the status becomes stale when waiting to be updated in the out-of-date regime with large status update intervals; on the other hand, the transmission reliability is high in this regime since the network load is low and hence collisions rarely happen. 

The effectiveness of dynamic auxiliary cost adaptation in SMART is shown in Fig. \ref{fig_thres}. The reference design is the status-unaware scheme with the optimal update interval. SMART is tested with different initial auxiliary cost values, as shown in the $x$-axis; note that for fixed auxiliary cost schemes, the initial value never changes afterwards. It is found that the reinforcement learning-based scheme can adapt to different network conditions, as represented by different numbers of vehicles, and obtain the best performance. However, a fixed auxiliary cost scheme without considering the networking aspect is insufficient to provide robust performance.

\section{Related Work}
\label{sec_relatedWork}
The study on AoI from a queuing theory perspective has attracted extensive attentions. Basically, this line of work can be categorized as status-unaware since only a time metric, i.e., AoI, is concerned. The pioneer work in \cite{kaul11,kaul12} analyzes the tradeoff between sampling latency and queuing delay theoretically, with applications in vehicular networks. Afterwards, many researchers have extended this result to e.g., scenarios with packet management \cite{costa16}, multi-class queuing systems \cite{huang15}, gamma distribution for the service time \cite{najm16}, controlling the status packet arrival process instead of assuming it is random \cite{sun17_tit}, and energy harvesting sources \cite{arafa18}. In addition, Ref. \cite{ino17} derives the stationary distribution of AoI under various queuing disciplines, and Ref. \cite{jiang19_isit} models the spatially correlated statuses as a random field. Fundamentally, such status-unaware schemes, aiming for AoI minimization, have to account for the sampling process to avoid congestion. In contrast, our proposed design is status-aware, such that the sampling can be as frequent as possible, but the proposed parallel transmitter and receiver are responsible for understanding the status and decide whether it is transmitted OTA or leveraging model predictions.

Status update in wireless networks also gains much traction recently. Most existing works focus on AoI optimization. Ref. \cite{kadota18,hsu18} have considered the wireless broadcast networks wherein scheduling decisions are centralized, and adopted the Whittle's index approach. The wireless multi-access scenario, and hence decentralized scheduling decisions are mode, is considered in \cite{jiang18_iot,jiang18_itc,jiang19_tcom,yates17,kosta18,ali19}. Building on existing MAC-layer protocol, e.g., CSMA or ALOHA, the access probability of backoff window size is optimized in \cite{yates17,kosta18,ali19}, on account of the fact that nodes may have different channel conditions, service rates and packet arrival rates. Ref. \cite{jiang18_itc,jiang19_tcom} adopts the Whittle's index approach and associates the access probability with the index. In \cite{jiang18_iot}, a round-robin scheduling policy is shown to be asymptotically optimal when the number of nodes is large. In addition, Ref. \cite{talak18,talak_mobihoc} show that a stationary policy actually achieves order-optimal performance in general network topology. SMART is inspired by the Whittle's index methodology, but generalizes to arbitrary status vectors (coped with by DNN) and not restricted to a certain network topology.

Another related line of work focuses on status-aware update schemes \cite{jiang19_wcm}. In \cite{sun17_wiener,orn19}, the authors propose a status-error threshold-based approach to optimally sample from Wiener and Ornstein-Uhlenbeck processes, respectively. Ref. \cite{kam18} proposes an effective AoI concept to better capture the status variation instead of simply AoI. In this regard, value of information \cite{kosta17,ayan19} also aims to capture the effect of AoI on specific application performance. For optimizations in wireless networks, a mean-field approach is utilized to calculate the access probabilities in \cite{jiang19_infocom} with random-walk status transitions. A simplified approach is to assign priorities to certain status packets considering their contents \cite{najm18_isit}. Most of the mentioned works have to assume a status model, e.g., Markov source model, for theoretical tractability, and have not considered status predictions. In contrast, this paper is the first to introduce status model predictions, i.e., parallel communications, and realize the concept on an SDR testbed.

\section{Conclusions and Outlook}
\label{sec_cl}
In this paper, we proposed a parallel communication scheme whereby status information is communicated by both OTA packets and aligned predictions by calibrated status models. The system is implemented on an SDR platform, and link-level experimental results show that the network load can be significantly reduced while maintaining low status recovery error by leveraging model predictions, namely revealing much while saying less. In addition, SMART is proposed for networking such predictive wireless devices in an ad hoc WNC system based on a Whittle's index-inspired reinforcement learning framework. To test the proposed approach in practice, we simulate on SUMO a multi dense platooning application with C-V2X communication protocol. It is shown by vehicle control performance that the parallel communication scheme significantly outperforms both AoI-optimized status-unaware schemes and conventional uRLLC schemes, due to the fact that the proposed scheme can reduce the network load, thus improving transmission reliability with less packet collisions.

Future research topics include investigating more robust system identification, model estimation and time-series forecasting mechanisms for status model predictions. More generally, the interplay between sensing, communication, computation and control is worth studying to enable more efficient wireless applications.

\section*{Acknowledgement}
This work was supported by the National Key R$\&$D Program of China (No. 2017YFE0121400), the program for Professor of Special Appointment (Eastern Scholar) at Shanghai Institutions of Higher Learning, and Shanghai Institute for Advanced Communication and Data Science (SICS).

\bibliographystyle{ieeetr}
\bibliography{aoi}

\begin{thebibliography}{10}

\bibitem{kaul12}
S.~Kaul, R.~Yates, and M.~Gruteser, ``Real-time status: How often should one
  update?,'' in {\em IEEE Conf. Comput. Commun. (INFOCOM)}, pp.~2731--2735, Mar
  2012.

\bibitem{sun17_tit}
Y.~{Sun}, E.~{Uysal-Biyikoglu}, R.~D. {Yates}, C.~E. {Koksal}, and N.~B.
  {Shroff}, ``Update or wait: How to keep your data fresh,'' {\em IEEE Trans.
  Inform. Theory}, vol.~63, pp.~7492--7508, Nov 2017.

\bibitem{usrp}
Online; \url{http://www.ni.com}; accessed 20-Jul-2019.

\bibitem{mol17}
R.~{Molina-Masegosa} and J.~{Gozalvez}, ``{LTE-V} for sidelink {5G V2X}
  vehicular communications: A new {5G} technology for short-range
  vehicle-to-everything communications,'' {\em {IEEE} Veh. Technol. Mag.},
  vol.~12, pp.~30--39, Dec 2017.

\bibitem{vuk18}
V.~Vukadinovic, K.~Bakowski, P.~Marsch, I.~D. Garcia, H.~Xu, M.~Sybis,
  P.~Sroka, K.~Wesolowski, D.~Lister, and I.~Thibault, ``{3GPP C-V2X} and
  {IEEE} 802.11p for vehicle-to-vehicle communications in highway platooning
  scenarios,'' {\em Ad Hoc Networks}, vol.~74, pp.~17 -- 29, 2018.

\bibitem{web90}
R.~R. Weber and G.~Weiss, ``On an index policy for restless bandits,'' {\em
  Journal of Applied Probability}, vol.~27, no.~3, p.~637–648, 1990.

\bibitem{pes00}
L.~Peshkin, K.-E. Kim, N.~Meuleau, and L.~P. Kaelbling, ``Learning to cooperate
  via policy search,'' in {\em Proceedings of the Conference on Uncertainty in
  Artificial Intelligence}, pp.~489--496, 2000.

\bibitem{kadota18}
I.~Kadota, A.~Sinha, E.~Uysal-Biyikoglu, R.~Singh, and E.~Modiano, ``Scheduling
  policies for minimizing age of information in broadcast wireless networks,''
  {\em IEEE/ACM Trans. Netw.}, vol.~26, pp.~2637--2650, Dec. 2018.

\bibitem{hsu18}
Y.~{Hsu}, ``Age of information: Whittle index for scheduling stochastic
  arrivals,'' in {\em IEEE Int'l Symp. Info. Theory}, pp.~2634--2638, Jun.
  2018.

\bibitem{jiang18_itc}
Z.~{Jiang}, B.~{Krishnamachari}, S.~{Zhou}, and Z.~{Niu}, ``Can decentralized
  status update achieve universally near-optimal age-of-information in wireless
  multiaccess channels?,'' in {\em International Teletraffic Congress (ITC
  30)}, vol.~01, pp.~144--152, Sep. 2018.

\bibitem{ddpg}
D.~Silver, G.~Lever, N.~Heess, T.~Degris, D.~Wierstra, and M.~Riedmiller,
  ``Deterministic policy gradient algorithms,'' in {\em International
  Conference on Machine Learning (ICML)}, pp.~387--395, 2014.

\bibitem{v2x}
Qualcomm, The path to 5G: Cellular Vehicle-to-Everything (C-V2X),
  https://www.qualcomm.com/documents/path-5g-cellular-vehicle-everything-c-v2x.

\bibitem{mode4}
3GPP TS 36.300: “Evolved universal terrestrial radio access (E-UTRA) and
  evolved universal terrestrial radio access network (E-UTRAN); overall
  description; stage 2”.

\bibitem{kaul11}
S.~Kaul, M.~Gruteser, V.~Rai, and J.~Kenney, ``Minimizing age of information in
  vehicular networks,'' in {\em IEEE Int. Conf. Sens., Commun., Netw. (SECON)},
  pp.~350--358, Jun 2011.

\bibitem{costa16}
M.~Costa, M.~Codreanu, and A.~Ephremides, ``On the age of information in status
  update systems with packet management,'' {\em IEEE Trans. Inform. Theory},
  vol.~62, pp.~1897--1910, April 2016.

\bibitem{huang15}
L.~Huang and E.~Modiano, ``Optimizing age-of-information in a multi-class
  queueing system,'' in {\em IEEE Int'l Symp. Info. Theory (ISIT)},
  pp.~1681--1685, Jun 2015.

\bibitem{najm16}
E.~Najm and R.~Nasser, ``The age of information: The gamma awakening,'' in {\em
  IEEE Int'l Symp. Info. Theory (ISIT)}, pp.~2574--2578, 2016.

\bibitem{arafa18}
A.~{Arafa}, J.~{Yang}, S.~{Ulukus}, and H.~V. {Poor}, ``Age-minimal online
  policies for energy harvesting sensors with incremental battery recharges,''
  in {\em Information Theory and Applications Workshop (ITA)}, pp.~1--10, Feb
  2018.

\bibitem{ino17}
Y.~{Inoue}, H.~{Masuyama}, T.~{Takine}, and T.~{Tanaka}, ``The stationary
  distribution of the age of information in {FCFS} single-server queues,'' in
  {\em IEEE Int'l Symp. Info. Theory (ISIT)}, pp.~571--575, June 2017.

\bibitem{jiang19_isit}
Z.~Jiang and S.~Zhou, ``Status from a random field: How densely should one
  update?,'' in {\em IEEE Int'l Symp. Info. Theory (ISIT)}, 2019.

\bibitem{jiang18_iot}
Z.~{Jiang}, B.~{Krishnamachari}, X.~{Zheng}, S.~{Zhou}, and Z.~{Niu}, ``Timely
  status update in wireless uplinks: Analytical solutions with asymptotic
  optimality,'' {\em IEEE Internet of Things Journal}, vol.~6, pp.~3885--3898,
  Apr 2019.

\bibitem{jiang19_tcom}
J.~Sun, Z.~Jiang, B.~Krishnamachari, S.~Zhou, and Z.~Niu, ``Closed-form
  {Whittle's} index-enabled random access for timely status update,'' {\em IEEE
  Trans. Commun.}, to appear.

\bibitem{yates17}
R.~D. Yates and S.~K. Kaul, ``Status updates over unreliable multiaccess
  channels,'' in {\em IEEE Int'l Symp. Info. Theory}, pp.~331--335, Jun 2017.

\bibitem{kosta18}
A.~Kosta, N.~Pappas, A.~Ephremides, and V.~Angelakis, ``Age of information
  performance of multiaccess strategies with packet management,'' {\em arXiv
  preprint arXiv:1812.09201}, 2018.

\bibitem{ali19}
A.~Maatouk, M.~Assaad, and A.~Ephremides, ``Minimizing the age of information
  in a {CSMA} environment,'' {\em arXiv preprint arXiv:1901.00481}, 2019.

\bibitem{talak18}
R.~{Talak}, S.~{Karaman}, and E.~{Modiano}, ``Distributed scheduling algorithms
  for optimizing information freshness in wireless networks,'' in {\em IEEE
  Int. Workshop Signal Process. Adv. Wireless Commun. (SPAWC)}, pp.~1--5, Jun
  2018.

\bibitem{talak_mobihoc}
R.~Talak, S.~Karaman, and E.~Modiano, ``Optimizing information freshness in
  wireless networks under general interference constraints,'' in {\em ACM Int.
  Symp. Mobile Ad Hoc Netw. Comput. (MobiHoc)}, pp.~61--70, 2018.

\bibitem{jiang19_wcm}
Z.~Jiang, S.~Fu, S.~Zhou, Z.~Niu, S.~Zhang, and S.~Xu, ``{AI}-assisted low
  information latency wireless networking,'' {\em IEEE Wireless Commun.}, to
  appear.

\bibitem{sun17_wiener}
Y.~Sun, Y.~Polyanskiy, and E.~Uysal-Biyikoglu, ``Remote estimation of the
  {Wiener} process over a channel with random delay,'' in {\em IEEE Int'l Symp.
  Info. Theory (ISIT)}, pp.~321--325, Jun 2017.

\bibitem{orn19}
T.~Z. Ornee and Y.~Sun, ``Sampling for remote estimation through queues: Age of
  information and beyond,'' {\em arXiv preprint arXiv:1902.03552}, 2019.

\bibitem{kam18}
C.~{Kam}, S.~{Kompella}, G.~D. {Nguyen}, J.~E. {Wieselthier}, and
  A.~{Ephremides}, ``Towards an effective age of information: Remote estimation
  of a markov source,'' in {\em IEEE Conference on Computer Communications
  Workshops (INFOCOM WKSHPS)}, pp.~367--372, Apr. 2018.

\bibitem{kosta17}
A.~Kosta, N.~Pappas, A.~Ephremides, and V.~Angelakis, ``Age and value of
  information: Non-linear age case,'' in {\em IEEE Int'l Symp. Info. Theory},
  pp.~326--330, Jun 2017.

\bibitem{ayan19}
O.~Ayan, M.~Vilgelm, M.~Kl\"{u}gel, S.~Hirche, and W.~Kellerer,
  ``Age-of-information vs. value-of-information scheduling for cellular
  networked control systems,'' in {\em ACM/IEEE International Conference on
  Cyber-Physical Systems}, pp.~109--117, 2019.

\bibitem{jiang19_infocom}
Z.~Jiang, S.~Zhou, Z.~Niu, and Y.~Cheng, ``A unified sampling and scheduling
  approach for status update in wireless multiaccess networks,'' in {\em IEEE
  Conf. Comput.Commun. (INFOCOM)}, May 2019.

\bibitem{najm18_isit}
E.~{Najm}, R.~{Nasser}, and E.~{Telatar}, ``Content based status updates,'' in
  {\em IEEE Int'l Symp. Info. Theory (ISIT)}, pp.~2266--2270, Jun 2018.

\end{thebibliography}
\end{document}